\documentstyle[12pt, a4]{article}

\begin{document}
\author{B. G\"{o}n\"{u}l, O. \"{O}zer, M. Ko\c{c}ak, D. Tutcu and Y. Can\c{c}elik
\and Department of Engineering Physics, University of Gaziantep, \and 27310
Gaziantep-T\"{u}rkiye}
\title{Supersymmetry and the relationship between a class of singular potentials in arbitrary dimensions}
\maketitle
\begin{abstract}
The eigenvalues of the potentials $V_{1}(r)=\frac{A_{1}}{r}+\frac{A_{2}}{%
r^{2}}+\frac{A_{3}}{r^{3}}+\frac{A_{4}}{r^{4}}$ and $V_{2}(r)=B_{1}r^{2}+\frac{%
B_{2}}{r^{2}}+\frac{B_{3}}{r^{4}}+\frac{B_{4}}{r^{6}}$, and of
the special cases of these potentials such as the Kratzer and
Goldman-Krivchenkov potentials, are obtained in $N$-dimensional
space. The explicit dependence of these potentials in
higher-dimensional space is discussed, which have not been
previously covered.
\end{abstract}
{\bf Pacs Numbers}:~03.65.Fd, 03.65.Ge

\section{Introduction}
Singular potentials have attracted much attention in recent years
for a variety of reasons, two of them being that (i) the ordinary
perturbation theory fails badly for such potentials, and (ii) in
physics, one often encounters phenomenological potentials that
are strongly singular at the origin such as certain type of
nucleon-nucleon potentials, singular models of fields in zero
dimensions, etc. Thus a study of such potentials is of interest,
both from the fundamental and applied point of view.

One of the challenging problems in non-relativistic quantum mechanics is to
find exact solutions to the Schr\"{o}dinger equation for potentials that can
be used in different field of physics. Recently, several authors obtained
exact solutions for the fourth-order inverse-power potential
\begin{equation}
\label{spot1}
V_{1}(r)=\frac{A_{1}}{r}+\frac{A_{2}}{r^{2}}+\frac{A_{3}}{r^{3}}+\frac{A_{4}}{r^{4}}
\end{equation}
using analytical methods [1-3]. These methods yield exact
solutions for a single state only for a potential of type
(\ref{spot1}) with restrictions on the coupling constants. The
interest is mainly due to the wide applicability of these type
inverse-power potentials. Some areas of interest are ion-atom
scattering \cite{gribakun}, several interactions between the atoms
\cite {vogt}, low-energy physics \cite{barut2}, interatomic
interactions in molecular physics \cite{kaplus} and solid-state
physics \cite{sherry}.

The advent of supersymmetry has had a significant impact on
theoretical physics in a number of distinct disciplines. One
subfield that has been receiving much attention is supersymmetric
quantum mechanics \cite{foran} in which the Hamiltonians of
distinct systems are related by a supersymmetry algebra. In this
work, we are concerned with, via supersymmetric quantum mechanics,
clarifying the relationship between two distinct systems having an
interaction potential of type (\ref{spot1}) and interacting
through
\begin{equation}
\label{spot2}
V_{2}(r)=B_{1}r^{2}+\frac{B_{2}}{r^{2}}+\frac{B_{3}}{r^{4}}+\frac{B_{4}}{r^{6}}
\end{equation}
singular even-power potentials which have been widely used in a
variety of fields, e.g. see [6,10]. In recent years, the higher
order anharmonic potentials have drawn more attentions of
physicists and mathematicians in order to partly understand a
newly discovered phenomena such as the structural phase
transitions \cite{share}, the polaron formation in solids
\cite{emin}, the concept of false vacuo in field theory
\cite{coleman}, fibre optics \cite{hashimoto}, and molecular
physics \cite{child}. In addition, some 60 years ago Michels {\it
et al.}. \cite{michels} proposed the idea of simulating the effect
of pressure on an atom by enclosing it in a impenetrable spherical
box. Since that time there have been a large number of
publications, for an overview see \cite{varshni}, dealing with
studies on quantum systems enclosed in boxes, which involve an
interaction potential that is a special case $(B_{2}=0)$ of
(\ref{spot2}). This field has received added impetus in recent
years because of the fabrication of semiconductor quantum dots
\cite{reed}.

The main motivation behind this work is to reveal the existence of
a link between potentials of type (\ref{spot1}) and (\ref{spot2})
in N-dimensional space, and between their special cases such as a
Mie-type potential (or Kratzer) \cite{mie} and pseudoharmonic-like
(or Goldman-Krivchenkov) potential \cite{goldman} in higher
dimensions, which to our knowledge has never been appeared in the
literature. On the other hand, with the advent of growth technique
for the realization of the semiconductor quantum wells, the
quantum mechanics of low-dimensional systems has become a major
research field. The work presented in this letter would also be
helpful to the literature in this respect as the results can
readily be extended to lower dimensions as well.

\section{The Schr\"{o}dinger equation in $N$-dimensional space}
It is well known that the general framework of the
non-relativistic quantum mechanics is by now well understood and
its predictions have been carefully proved against observations.
Physics is permanently developing in a tight interplay with
mathematics. It is of importance to know therefore whether some
familiar problems are a particular case of a more general scheme
or to search if a map between the radial equations of two
different systems exists. It is hence worthwhile to study the
Schr\"{o}dinger equation in the arbitrary dimensional spaces
which has attracted much more attention to many authors. Many
efforts have in particular been produced in the literature over
several decades to study the stationary Schr\"{o}dinger equation
in various dimensions with a central potential containing
negative powers of the radial coordinates [21, and the
references therein].

The radial Schr\"{o}dinger equation for a spherically symmetric potential in
N-dimensional space (we shall use through this paper the natural units such
that $\hbar=m=1)$
\begin{equation}
\label{sschr1}
-\frac{1}{2}\left[\frac{d^{2}R}{dr^{2}}+\frac{N-1}{r}\frac{dR}{dr}\right]+\frac{%
\ell(\ell+N-2)}{2r^{2}}R =[E-V(r)]R
\end{equation}
is transformed to
\begin{equation}
\label{strans1}
-\frac{d^{2}\Psi}{dr^{2}}+\left[\frac{(M-1)(M-3)}{4r^{2}}+2V(r)\right]
\Psi=2E\Psi
\end{equation}
where $\Psi$, the reduced radial wave function, is defined by
\begin{equation}
\label{swave1}
\Psi(r)=r^{(N-1)/2}R(r)
\end{equation}
and
\begin{equation}
\label{sm1}
M=N+2\ell
\end{equation}
Eq.(\ref{strans1}) can also be written as
\begin{equation}
\label{sschr2}
-\frac{1}{2}\frac{d^{2}\Psi}{dr^{2}}+\left[\frac{\Lambda(\Lambda+1)}{2r^{2}}%
+V(r)\right]\Psi=E\Psi
\end{equation}
where $\Lambda=(M-3)/2$. We see that the radial Schr\"{o}dinger
equation in $N$-dimensions has the same form as the
three-dimensional one. Consequently, given that the potential has
the same form in any dimension, the solution in three dimensions
can be used to obtain the solution in any dimension simply by
using the substitution $\ell\rightarrow\Lambda$. It should be
noted that $N$ and $\ell$ enter into expressions (\ref{strans1})
and (\ref{sschr2}) in the form of the combinations $N+2\ell$.
Hence, the solutions for a particular central potential $V(r)$ are
the same as long as $M(=N+2\ell)$ remains unaltered. Therefore the
s-wave eigensolutions $(\Psi_{\ell=0})$ and eigenvalues in
four-dimensional space are identical to the p-wave solutions $%
(\Psi_{\ell=1}) $ in two-dimensions.

The technique of changing the independent coordinate has always been useful
tool in the solution of the Schr\"{o}dinger equation. For instance, this
allows something of a systematic approach enabling to recognize the
equivalence of superficially unrelated quantum mechanical problems. Many
recent papers have addressed this old subject. In the light of these works
we proceed by substituting $r=\alpha\rho^{2}/2$ and $R=F(\rho)/\rho^{%
\lambda} $, $\lambda$ an integer, suggested by the known
transformations between Coulomb and harmonic oscillator problems
\cite{schrodinger} and used to show the relation between the
perturbed Coulomb problem and the sextic anharmonic oscillator in
arbitrary dimensions [23,24], we transform Eq. (\ref{sschr1}) to
another Schr\"{o}dinger-like equation in $N^{\prime}=2N-2-2\lambda
$ dimensional space with angular momentum $L=2\ell+\lambda$,
\begin{equation}
\label{sschr3}
-\frac{1}{2}\left[\frac{d^{2}F}{d\rho^{2}}+\frac{N^{\prime}-1}{\rho}\frac{dF}{%
d\rho}\right]+\frac{L(L+N^{\prime}-2)}{2\rho^{2}}F =[\hat{E}-\hat{V}(\rho)]F
\end{equation}
where
\begin{equation}
\label{sequal1}
\hat{E}-\hat{V}(\rho)=E\alpha^{2}\rho^{2}-\alpha^{2}\rho^{2}V(\alpha%
\rho^{2}/2)
\end{equation}
and $\alpha$ is a parameter to be adjusted properly. Note that leaving
re-scaling constant $\alpha$ arbitrary for now gives us an additional degree
of freedom. When we discuss bound state eigenvalues later, we can use this
to allow the values of the potential coefficients to be completely
independent of each other. Thus, by this transformation, in general, the $N$%
-dimensional radial wave Schr\"{o}dinger equation with angular momentum $%
\ell $ can be transformed to a $N^{\prime}=2N-2-2\lambda$ dimensional
equation with angular momentum $L=2\ell+\lambda$. If we choose $%
\alpha^{2}=1/|E|$, with $E$ corresponding the eigenvalue for the
inverse power potential of Eq. (\ref{spot1}), then Eq.
(\ref{sschr3}) corresponds to the Schr\"{o}dinger equation of an
singular even-power potential
\begin{equation}
\label{stranspot1}
\hat{V}(\rho)=\rho^{2}+\frac{4A_{2}}{\rho^{2}}+
\frac{8A_{3}}{\rho^{4}}|E|^{1/2} +\frac{16A_{4}}{\rho^{6}}|E|
\end{equation}
with eigenvalue
\begin{equation}
\label{stransenergy1}
\hat{E}=\frac{-2A_{1}}{|E|^{1/2}}
\end{equation}

Thus, the system given by Eq. (\ref{spot1}) in $N$-dimensional
space is reduced to another system defined by Eq. (\ref{spot2}) in
$N^{\prime}=2N-2-2\lambda$ dimensional space. In other words, by
changing the independent variable in the radial Schr\"{o}dinger
equation, we have been able to demonstrate a close
equivalence between singular potentials of type (\ref{spot1}) and (\ref{spot2}). Note that when $%
N=3$ and $\lambda=0$ one finds $N^{\prime}=4$, and when $\lambda=1$ we get $%
N^{\prime}=2$. It is also easy to see that $N^{\prime}+2L$ does not depend
on $\lambda$, which leads to map two distinct problems in three- and
four-dimensional space \cite{morales}.

\section{Mappings between two distinct systems}

\subsection{Quasi-exactly solvable case}
Since Eq. (\ref{strans1}) for the reduced radial wave $\Psi(r)$ in
the $N$-dimensional space has the structure of the one-dimensional
Schr\"{o}dinger equation for a spherically symmetric potential
$V(r)$, we may define the supersymmetric partner potentials
\cite{foran}
\begin{equation}
\label{spartners}
V_{\pm}(r)=W^{2}(r)\pm W^{\prime}(r)
\end{equation}
which has a zero-energy solution, and the corresponding eigenfunction is
given by
\begin{equation}
\label{swave2}
\Psi_{n=0}(r)\propto exp[\pm \int^{r}W(r) dr]
\end{equation}

In constructing these potentials one should be careful about the
behaviour of the wave function $\Psi(r)$ near $r=0$ and
$r\rightarrow \infty$. It may be mentioned that $\Psi(r)$ behaves
like $r^{(M-1)/2}$ near $r=0$ and it should be normalizable. For
the inverse power potential of Eq. (\ref{spot1}) we set
\begin{equation}
\label{ssuper1}
W(r)=\frac{-a}{r^{2}}+\frac{c}{r}-b~~~,~~~b,c>0
\end{equation}
and identify $V_{+}(r)$with the effective potential so that
\begin{equation}
\label{spartpot1}
V_{+}(r)=\left(\frac{2A_{4}}{r^{4}}+\frac{2A_{3}}{r^{3}}+\frac{2A_{2}}{r^{2}}+%
\frac{2A_{1}}{r}\right)+\frac{(M-1)(M-3)}{4r^{2}}-2E_{n=0}
\end{equation}
and substituting Eq. (\ref{ssuper1}) into Eq. (\ref{spartners}) we
obtain
\begin{equation}
\label{spartpot2}
V_{+}(r)=\frac{a^{2}}{r^{4}}+\frac{2a(1-c)}{r^{3}}+\frac{c(c-1)+2ab}{r^{2}}-%
\frac{2bc}{r}+b^{2}
\end{equation}
and the relations between the parameters satisfy the supersymmetric
constraints
\begin{equation}
\label{sparameter1}
a=\pm\sqrt{2A_{4}}~~~ ;~~~
c=1-\frac{A_{3}}{\pm\sqrt{2A_{4}}}
\end{equation}
The potential (\ref{spot1}) admits the exact solutions
\begin{equation}
\label{swave3}
\Psi_{n=0}(r)= N_{0}~r^{c}~exp(\frac{a}{r}-br)
\end{equation}
where $N_{0}$ is the normalization constant, with the physically acceptable
eigenvalues
\begin{equation}
\label{senergy1}
E_{n=0}=-\frac{b^{2}}{2}=-\frac{1}{16A_{4}}\left[\frac{A_{3}}{\sqrt{2A_{4}}}
(1+\frac{A_{3}}{\sqrt{2A_{4}}})-\frac{1}{4}(M-1)(M-3)-2A_{2}\right]^{2}
\end{equation}
in the case of $a<0$ and under the constraints
\begin{equation}
\label{sa1}
A_{1}=-(1+\frac{A_{3}}{\sqrt{2A_{4}}})\sqrt{-2E_{n=0}}
\end{equation}
The results obtained agree with those in Refs. [2,3,21] for
three-dimensions. Note that in order to retain the well-behaved solution at $%
r\rightarrow 0$ and at $r\rightarrow\infty$ we have chosen
$a=-\sqrt{2A_{4}}$.

The expressions obtained above can easily be extended to the lower
dimensions. For example, one can readily check that our two-dimensional
solutions ($N=2,\ell\rightarrow\ell-1/2$) for the inverse power potential
considered are in excellent agreement with the literature \cite{dong}. The
ground state solutions in arbitrary dimensions for the Coulomb ($%
A_{2}=A_{3}=A_{4}=0$), and for a the Kratzer $(A_{3}=A_{4}=0)$ \cite{mie},
and for an inverse-power $A_{3}=0$ [1,2] potentials can also be found from
the above prescriptions.

For the singular even-power anharmonic oscillator potential of Eq.
(\ref{spot2}), we set
\begin{equation}
\label{ssuper2}
W(r)=\mu r+ \frac{\delta}{r} +
\frac{\eta}{r^{3}}~~,~~\delta>0
\end{equation}
which leads to
\begin{equation}
\label{swave4}
\Psi_{n=0}(r)= C_{0}~r^{\delta}~ exp(\frac{\mu
r^{2}}{2}-\frac{\eta}{2r^{2}})
\end{equation}
with $C_{0}$ being the corresponding normalization constant, and identify $%
V_{+}(r)$ with the effective potential so that
\begin{eqnarray}
V_{+}(r)&=&\left(\frac{2B_{4}}{r^{6}}+\frac{2B_{3}}{r^{4}}+\frac{2B_{2}}{r^{2}}%
+2B_{1}r^{2}\right)+\frac{(M-1)(M-3)}{4r^{2}}-2\tilde{E}_{n=0}  \nonumber
\end{eqnarray}
\begin{equation}
\label{spartpot3}
=W^{2}(r)+W^{\prime }(r)=\frac{\eta ^{2}}{r^{6}}+\frac{\eta (2\delta -3)}{%
r^{4}}+\frac{\delta (\delta -1)+2\eta \mu }{r^{2}}+\mu ^{2}r^{2}+\mu
(2\delta +1)
\end{equation}
and the relations between the potential parameters satisfy the
supersymmetric constraints
\begin{equation}
\label{seta1}
\eta=\pm\sqrt{2B_{4}}~~;~~\delta=\frac{3}{2}+\frac{B_{3}}{%
\eta}~~;~~\mu=\pm\sqrt{2B_{1}}
\end{equation}
As we are dealing with a confined particle system, the positive values for $%
\eta$ and the negative values for $\mu$ would of course be the right choice to ensure the well
behaved nature of the wave function behaviour at the origin and at infinity.
Hence, physically meaningful ground state energy eigenvalues for the
potential of interest are
\begin{equation}
\label{stransenergy2}
\tilde{E}_{n=0}=-\frac{\mu}{2}(2\delta+1)=\sqrt{\frac{B_{1}}{2}}\left\{2+\sqrt{%
1-16\sqrt{B_{1}B_{4}}+ 8B_{2}+(M-1)(M-3)}\right\}
\end{equation}
At this point we should report that our results reproduce those
obtained by [17,25,26] when potential (\ref{spot2}) (in case
$B_{2}=0$) is confined to an impenetrable spherical box in 2- and
3-dimensions. It is also not difficult to see that if one takes
$\eta=0$ in Eq. (\ref{spartpot3}), then Eq. (\ref{stransenergy2})
becomes the exact energy spectra of $N$-dimensional harmonic
oscillator. Further, one easily check that in case
$B_{4}=B_{3}=0$, the above energy expression correctly reproduce
the eigenvalues of the pseudo-type potential in 3-dimension
\cite{kasap} which is the subject of the next section.

Finally, we wish to discuss briefly the explicit mapping between
the singular potentials given by Eqs. (\ref{spot1}) and
(\ref{spot2}). If one consider the transformed anharmonic
oscillator potential of Eq. (\ref{stranspot1}) and repeat the
above mathematical procedure carried out through Eqs. (21-25),
then the corresponding eigenvalue equation reads
\begin{equation}
\label{stransenergy3}
\hat{E}_{n=0}=-2\mu
(1+\frac{A_{3}}{\sqrt{2A_{4}}})
\end{equation}
Using the physically acceptable definition of $A_{1}$ in Eq.
(\ref{sa1}), the above equation can be rearranged as
\begin{equation}
\label{stransenergy4}
\hat{E}_{n=0}=-\frac{2A_{1}}{|E_{n=0}|^{1/2}}
\end{equation}
where $E_{n=0}$ has been described in Eq. (\ref{senergy1}). This
brief discussion shows explicitly the relation between the two
singular potentials in higher dimensions and verifies Eq.
(\ref{stransenergy1}).

\subsection{Exactly solvable case}
Kasap \cite{kasap} and his co-workers used supersymmetric quantum
mechanics to find exact results for the special cases of the
singular potentials of (\ref{spot1}) and (\ref{spot2}), more
precisely the solutions of the Kratzer and pseudoharmonic
potentials in three dimensions. Their results can be easily
generalized to $N$-dimensions by the substitution
$\ell\rightarrow\Lambda=(M-3)/2$ as indicated in section II. This
extension to arbitrary dimensions helps us in constructing the map
between these two distinct systems.

The study of anharmonic oscillators has raised a considerable amount of
interest because of its various applications especially in molecular
physics. The Morse potential is commonly used for anharmonic oscillator.
However, its wave function does not vanish at the origin, but those for
Mie-type and pseudoharmonic potentials do. The Mie-type potential possesses
the general features of the true interaction energy, inter-atomic and
inter-molecular, and dynamical properties of solids \cite{maitland}. On the
other hand, the pseudoharmonic potential may be used for the energy spectrum
of linear and non-linear systems \cite{goldman}. The Mie-type and
pseudo-harmonic potentials are two special kinds of analytically solvable
singular-power potentials as they have the property of shape-invariance.

Starting with the general form of the Mie-type potential
\begin{equation}
\label{smiepot1}
V(r)=D_{0}\left[\frac{p}{q-p}(\frac{\sigma}{r})^{q}-\frac{q}{q-p}(\frac{\sigma}{r}%
)^{p}\right]
\end{equation}
where $D_{0}$ is the interaction energy between two atoms in a molecular
system at $r=\sigma$, and $q>p$ is always satisfied. If we take $q=2p$ and $%
p=1$, we arrive at a special case of the potential in Eq.
(\ref{smiepot1}), which is exactly solvable
\begin{equation}
\label{smiepot2}
V(r)=\frac{A}{r^{2}}-\frac{B}{r}
\end{equation}
where $A=D_{0}\sigma^{2}$ and $B=2D_{0}\sigma$. The above
potential, the so-called Kratzer potential, includes the terms
which give the representation of both the steep repulsive branch
and the long-range
attraction. A single minimum occurs at $r=\sigma$ where the energy is $%
-D_{0} $. Considerable interest has recently been shown in this potential as
a model to describe inter-nucleon vibration \cite{secrets} and, in
applications this Mie type potential offers one of the most important
exactly solvable models of atomic and molecular physics and quantum
chemistry \cite{znojil}.

We set the superpotential for the Kratzer effective potential
\begin{equation}
\label{ssuper3}
W(r)=\frac{B/2}{\beta+(\beta^{2}+C)^{1/2}}-\frac{\beta+(\beta^{2}+C)^{1/2}}{r}
\end{equation}
where
\begin{equation}
\label{sc}
C=\frac{\Lambda(\Lambda+1)}{2}+A~~,~~\Lambda=\ell+\frac{1}{2%
}(N-3)~~,~~\beta=\frac{1}{2\sqrt{2}}
\end{equation}
and obtained the exact spectrum in $N$-dimensional space as
\begin{equation}
\label{senergy2}
E_{n}=\left(\frac{B/2\beta}{2n+1+[(2\Lambda+1)^{2}+A/\beta^{2}]^{1/2}}\right)^{2}
~~,~~n=0,1,2,...
\end{equation}
and from Eq. (\ref{swave2}) the exact unnormalized ground state
wavefunction can be expressed as
\begin{equation}
\label{swave5}
\Psi_{n=0}(r)=r^{1/2\{1+[(2\Lambda+1)^{2}+A/\beta^{2}]^{1/2}}\} \times exp(-%
\frac{Br/4\beta^{2}}{1+[(2\Lambda+1)^{2}+A/\beta^{2}]^{1/2}})
\end{equation}
The excited state wavefunctions can be easily determined from the usual
approach in supersymmetric quantum mechanics \cite{foran} and the
normalization coefficients for each quantum state wave function can be
analytically worked out using the explicit recurrence relation given in a
recent work \cite{jia}.

As a second application, we consider the general form of the pseudoharmonic
potential
\begin{equation}
\label{stranspot2}
\tilde{V}(r)=V_{0}(\frac{r}{r_{0}}-\frac{r_{0}}{r})^{2}=\tilde{B}r^{2} +%
\frac{\tilde{A}}{r^{2}}-2V_{0}
\end{equation}
which can be used to calculate the vibrational energies of diatomic
molecules with the equilibrium bond length $r_{0}$ and force constant $%
k=8V_{0}/r_{0}^{2}$, and set the corresponding superpotential as
\begin{equation}
\label{ssuper4}
W(r)=\sqrt{\tilde{B}}r-\frac{\beta+(\beta^{2}+\tilde{C})^{1/2}}{r}
\end{equation}
where $\tilde{B}=V_{0}/r_{0}^{2}$, $\tilde{C}=[\Lambda(\Lambda+1)+2\tilde{A}%
]/2$, $\tilde{A}=V_{0}r_{0}^{2}$. The exact full spectrum of the potential
in arbitrary dimensions is
\begin{equation}
\label{stransenergy5}
\tilde{E_{n}}=2\beta\sqrt{\tilde{B}}\{4n+2+[(2\Lambda+1)^{2}+\tilde{A}%
/\beta^{2}]^{1/2}\}-2V_{0}
\end{equation}
and the unnormalized exact ground state wave function is
\begin{equation}
\label{swave6}
\Psi_{n=0}(r)=r^{1/2\{1+[(2\Lambda+1)^{2}+\tilde{A}/\beta^{2}]^{1/2}}\}
\times exp(-\frac{\sqrt{\tilde{B}}r^{2}}{4\beta})
\end{equation}

Using the discussion in section 2, one can transform the Kratzer
potential in Eq. (\ref{smiepot2}) to its dual potential- shifted
(by $2V_{0}$) pseudoharmonic-like potential in Eq.
(\ref{stranspot2}) with some restrictions in potential parameters.
In the light of Eqs. (9-11), the transformed potential reads
\begin{equation}
\label{stranspot3}
\hat{V}(\rho)=\rho^{2}+\frac{4A}{\rho^{2}}
\end{equation}
which is in the form of the Goldman-Krivchenkov potential. Here $%
A(=D_{0}\sigma^{2})$is the Kratzer potential parameter and, considering Eqs.
(\ref{stranspot2}) through Eq. 36, constraints on the potential parameters are such that $%
\tilde{B}=1$ and $\tilde{A}=4A$. In this case corresponding eigenvalues are
\begin{eqnarray}
\label{stransenergy6}
\hat{E}_{n^{\prime }}
&=&\frac{2B}{|E_{n}|^{1/2}}=4\beta \{1+2n^{\prime
}+\lbrack 1+4\Lambda ^{\prime }(\Lambda ^{\prime }+1)+\frac{A}{\beta ^{2}}%
\rbrack \}~~,~~ \nonumber \\
\Lambda ^{\prime } &=&L+\frac{1}{2}(N^{\prime }-3)
\end{eqnarray}
where $B(=2D_{0}\sigma )$ and $E_{n}$ are the coupling parameter and the
eigenenergy values (Eq. 32), respectively, of the Kratzer potential.

The ensuing relationships among the dimensionalities and quantum numbers of
the two distinct systems considered here in this section are :
\begin{equation}
\label{snprime}
N^{\prime}=2N-2-2\lambda~~,~~L=2\ell+\lambda~~~~
n^{\prime}=2n-2+\lambda
\end{equation}
Clearly, the mapping parameter $\lambda$ must be an integer if
$n^{\prime}$, $L$, $n$ and $\ell$ are integers. It is worthwhile
to discuss briefly the physics behind this transformation in the
light of the comprehensive work of Kostelecky {\it et al.}.
\cite{schrodinger}. We note that it is a general feature of this
map that the spectrum of the $N$-dimensional problem
involving Kratzer potential is related to the half the spectrum of the $%
N^{\prime}$-dimensional problem involving Goldman-Krivchenkov
potential for any even integer $N^{\prime}$. However, the
quantities in Eq. (\ref{snprime}) have parameter spaces that are
further restricted by the properties chosen for
the map. For instance, suppose we wish to map all states corresponding the $%
N $-dimensional Kratzer potential into those corresponding
Goldman-Krivchenkov potential. Since on physical grounds we know that $%
N^{\prime}\geq 2$, $n^{\prime}\geq 0$, $L\geq 0$, we must impose
$N\geq 2+\lambda$, $n\geq 1-\lambda/2$, $\ell\geq -\lambda/2$. This yields the bound $%
-2\ell\leq \lambda\leq N-2$. Further requiring $n\geq 1$, $\ell\geq 0$
restricts the bound to $0\leq \lambda\leq N-2$. We conclude that all states of the $N$%
-dimensional Kratzer problem can be mapped into the appropriate
Goldman-Krivchenkov problem, except for $N=1$.

As an example, consider the three-dimensional Kratzer problem.
Assuming we wish to map all its states into those of its dual-the
Goldman-Krivchenkov potential, we must impose $0\leq \lambda\leq 1$.
First, take $\lambda=0$. Then, the $s$-orbitals in Kratzer
potential ($n\geq 1,\ell=0$) are related to the
($n^{\prime}=2n-2\geq 0,L=0$ ) states of the four-dimensional
Goldman-Krivchenkov problem. Similarly, the $p$-states
($n\geq 2,\ell=1$)
correspond to the ($n^{\prime}=2n-2\geq 0,L=0$) same problem. Next, suppose $%
\Lambda=1$. The states corresponding the potential in Eq.
(\ref{smiepot2}) are then mapped into the odd-integer states of
the two-dimensional oscillator problem of Eq. (\ref{stranspot3}).
The $s$-orbitals of Kratzer potential ($n\geq1,\ell=0$) map into
the ($n^{\prime}=2n-1\geq1,L=1$) anharmonic oscillator states
corresponding Goldman-Krivchenkov potential, while the Kratzer $p$-orbitals (%
$n\geq2,\ell=1$) map into the ($n^{\prime}=2n-1\geq3,L=1$)
oscillator states of Goldman-Krivchenkov problem. As a rule, in
both cases ($\lambda=0,1$), the lowest-lying states of
Goldman-Krivchenkov potential are excluded , one by one, with
each higher value of $\ell$.

As a final remark, a student of introductory quantum mechanics
often learns that the Schr\"{o}dinger equation is exactly solvable
(for all angular momenta) for two central potentials in Eqs.
(\ref{smiepot2}) and (\ref{stranspot3}), and for also their
special cases ($A=0$) the Coulomb and harmonic oscillator
problems. Less frequently, the student made aware of the relation
between these two problems, which are linked by a simple change of
the independent variable discussed in detail through the paper.
Under this transformation, energies and coupling constants trade
places, and orbital angular momenta are rescaled. Thus, we have in
this section shown that there is really only one quantum
mechanical problem, not two involving the Kratzer and
Goldman-Krivchenkov potentials, which can be exactly solved for
all orbital angular momenta.

\section{Conclusion}
The main aim of this work has been to establish a very general
connection between a class of singular potentials in higher
dimensional space through the application of a suitable
transformation. Although much work had been done in the literature
on similar problems, an investigation as the one we have discussed
in this paper was missing to our knowledge. In addition, it is
shown that the supersymmetric quantum mechanics yields exact
solutions for a single state only for the quasi-exactly solvable
potentials such as the ones given in Eqs. (\ref{spot1}) and
(\ref{spot2}) with some restrictions on the potential parameters
in $N$-dimensional space, unlike the shape invariant exactly
solvable potentials. We have also shown how to obtain exact
solutions to such problems in any dimension by applying an
adequate transformation to previously known three-dimensional
results. This simple and intuitive method discussed through this
paper is easy to be generalized. The application of this method to
other potentials involving non-central ones are in progress.

\hspace{1.0in}

\newpage\

\end{document}